# LIPSS-Sticks: Laser induced double self organization enhances the broadband light harvesting of TiO$_2$ nanotube arrays

*Rakesh Arul[1-6]†, Junzhe Dong[6], M. Cather Simpson[1-5]\*, Wei Gao[6]\**

1. The Photon Factory, the University of Auckland
2. The MacDiarmid Institute for Advanced Materials and Nanotechonology
3. The Dodd Walls Centre for Quantum and Photonic Technologies
4. School of Chemical Sciences, the University of Auckland
5. Department of Physics, the University of Auckland
6. Department of Chemical and Materials Engineering, the University of Auckland

ABSTRACT: Sub-wavelength laser induced periodic surface structures (LIPSS-Sticks) created by ultrashort pulsed laser irradiation on the surface of titanium are used for the first time to template the electrochemical growth of titanium dioxide nanotube arrays. This is an example of a double self-organized process, as both LIPSS formation and electrochemical anodization involve spontaneous generation of order from initially non-ordered precursors. LIPSS-Sticks have a 2x greater visible to near infrared light (400 - 1400 nm) collection efficiency compared to flat titanium dioxide due to the enhanced light scattering from grating-like structures. The growth of nanostructures with time was modelled electrostatically to explain the features of a templated anodization process that differ from the usual anodization of flat surfaces. This new templated growth method is general and can also be applied to Cu, W, Fe, Ti alloys and Al for the fabrication of hierarchically nanostructured surfaces using two complementary fabrication techniques: ultrashort pulsed laser ablation and electrochemical anodization.

**Introduction**

In the pursuit of a sustainable future in the energy, transport and chemical industries, light-based technologies are leading the chase. The ability to store solar energy as fuel (e.g. hydrogen or methanol), to create chemical feedstock (e.g. nitrogen fixation [1]), or to remediate environmental pollutants through photocatalytic degradation holds the key to a "green" future. The first step in achieving these goals is the creation of a molecule/material that can efficiently absorb sunlight and turn solar energy into work in chemical reactions or electricity. The front-runner in the race for efficient photocatalysts are visible light-absorbing semiconductors. Semiconductors can absorb light with energy above its bandgap to create electron-hole pairs, which can then be used to perform redox reactions or generate electrical power. We use the classic transition metal oxide photocatalyst: TiO$_2$. Titanium dioxide has a long history in decomposing environmental pollutants [2] and water-splitting to form hydrogen [3, 4], but suffers from poor visible light absorption as it is a wide-bandgap semiconductor. In photovoltaics, $TiO_2$ is used as a support for dye-sensitized and metal-organic perovskite photovoltaics [5, 6] or dye-sensitized solar cells [7].

There exist two main paradigms to boost light collection efficiency and charge separation: (1) bandgap engineering through doping [8], alloying [9], defect control [10], lattice strain, and heterostructuring [11]; and (2) morphology engineering to reduce recombination, increase crystallinity, and harvest light more effectively. Approaches like thermal annealing introduce oxygen vacancies and increase visible light absorption [10, 12], but if too many defects are introduced, the electron-hole separation lifetime and the useful work extracted decreases, thus reducing overall photocatalytic reaction rates and efficiencies [13]. Here we attempt to circumvent this difficulty by nano-structuring the surface of the catalyst to more effectively harvest visible light while maintaining the elemental composition of the semiconductor. By multiple scattering of light in a nanostructured substrate, we increase the effective path length of light in the material, enabling a greater absorption of light even if the absorption coefficient of a material is small. This approach is novel and capable of being applied across a large portion of the transition metal oxide family, which spans bandgaps and redox potentials throughout a wide range of desired chemical reactions.

High aspect ratio metal oxide nanotubes can be produced using electrochemical anodization of metallic sheets [14]; and nanotube size, wall thickness, and array symmetry can be controlled by the anode voltage, anodization time, and electrolyte. This self-organized process creates hexagonally periodic nanotube arrays by balancing two reactions: metal oxidation to form an oxide and chemical dissolution of the oxide [15]. Indeed, the confluence of all these microscopic effects produces a large scale organized structure which is a paradigm of self-organization [16], and the true description is still an active area of research [17, 18]. Similarly, laser induced periodic surface structures (LIPSS) [19] are sub-wavelength self-organized structures that can be patterned by an ultrashort pulsed laser on a range of different metals, polymers and dielectrics [20]. LIPSS surface ripples have a periodicity that is smaller than the wavelength of the laser, with an orientation depending on the polarization [21, 22]. The sub-wavelength periodicity of LIPSS formation on metallic surfaces [23] is due to the interference of the incident light from the femtosecond pulse and a surface scattered wave [21] which creates a local intensity modulation that results in a ripple morphology [24]. However, this is not the sole formation mechanisms and competing theories have been proposed [25].

By patterning the surface of Ti metal with LIPSS prior to electrochemical anodization (Figure 1), we fabricated arrays of $TiO_2$ nanotubes that organized into ripples, thus acting as a grating to enhance the trapping of visible light [26-28]. The ability to pattern LIPSS on Ti has been previously reported [29-33] (Supplementary Table S1). However, we show below the first known case of LIPSS being used to template the growth of anodic $TiO_2$ nanotubes in a grating structure, and as a templating technique for nanostructured metal oxide growth in general, which we term "LIPSS-Sticks".

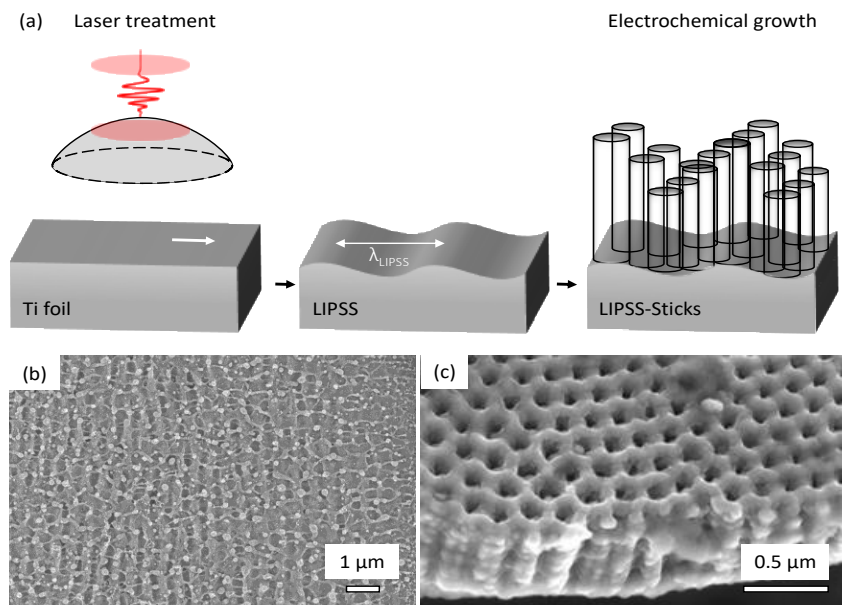

**Figure 1.** Double self-organization: (a) Schematic of self-organized laser induced periodic surface structures (LIPSS) and subsequent self-organized growth of nanotubes. Femtosecond pulsed laser irradiation is focused through a lens to treat the surface of Ti metal foil and create self-organized surface ripples (LIPSS) which are used as templates for self-organized electrochemical growth of $TiO_2$ nanotube arrays. SEM images of (b) LIPSS structure on titanium metal substrate, and (c) $TiO_2$ nanotube arrays electrochemically grown on flat/non laser-treated titanium metal.

## Results and Discussion

**Increased visible to near-infrared light harvesting efficiency of LIPSS-Sticks**

The laser fluence and number of overlapped laser pulses was varied (details in Supplementary Info S1) to determine the periodicity of LIPSS produced with femtosecond pulsed laser treatment (800 nm, 1 kHz, 130 fs) on flat electropolished Ti metal foil. The morphology of the LIPSS was imaged with an SEM and the 2D Fast Fourier Transform was used to determine the periodicity of the lines (Supplementary Info S2). We reproducibly fabricated sub-wavelength structures with periodicity between 250 to 650 nm. The laser conditions corresponding to a periodicity of 250 nm and 500 nm (105 mJ·cm$^{-2}$, 250 pulses; and 87 mJ·cm$^{-2}$, 50 pulses respectively) were chosen for further templating experiments and large samples (2 cm × 2 cm) were made for optical measurement.

Figure 2 shows the growth of the LIPSS structured surface under electrochemical anodization with time. Initially, the laser-treated surface showed only grating-like ripples from the LIPSS. After three hours of anodization, $TiO_2$ nanotubes with initial periodicity imprinted by the laser treatment were formed. After 7 hours, the rippled surface structure was destroyed as the nanotubes grew too long and were etched away, collapsing under their own weight [34]. Further details on the nanotube morphology evolution in a structured environment is discussed in a later section.

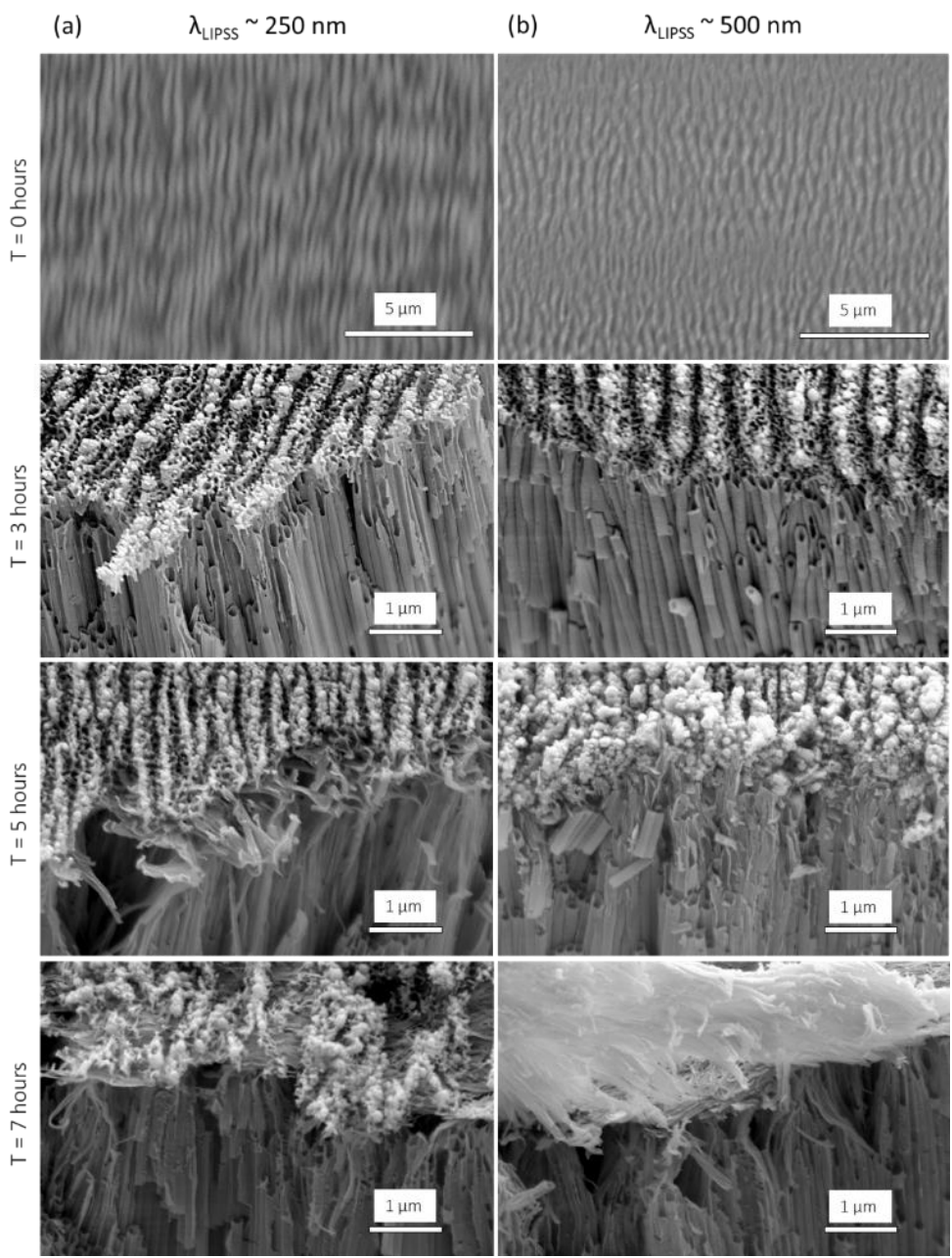

**Figure 2.** LIPSS-Sticks hierarchical nanostructure composed of sub-micron ripples on oxide nanotubes. Evolution with time over 7 hours is shown for electrochemical anodization of samples with initial LIPSS periods: (a) $\lambda_{LIPSS}$ = 250 nm and (b) $\lambda_{LIPSS}$ = 500 nm.

The LIPSS-Stick substrates for $\lambda_{LIPSS}$ = 250 nm and $\lambda_{LIPSS}$ = 500 nm initial LIPSS periodicities showed increased light extinction relative to the flat non-laser processed TiO$_2$ nanotube array (Figure 3a). Flat TiO2 already shows an appreciable visible-NIR light absorption beyond its band edge due to the scattering of the nanostructures on the film that is approximately 10 µm thick. Light extinction is increased over a large wavelength range, from the visible (near the band-edge of TiO$_2$) to the near-infrared (deep within the bandgap energy of TiO$_2$). As the chemical structure of TiO$_2$ nanotubes are very similar with and without LIPSS templating (measured by X-ray photoelectron spectroscopy and energy dispersive X-ray spectroscopy in Supplementary Info S4 and S5), the increased extinction must be due to the increased light scattering of the hierarchical LIPSS-Sticks structure, as opposed to an increased density of sub-bandgap defect states [10]. The nanostructured material scatters the light within and thereby increases the interaction between light and the material. If 40% of incident light is absorbed by the non-structured film, then increasing the number of scattering interfaces means more light is absorbed overall.

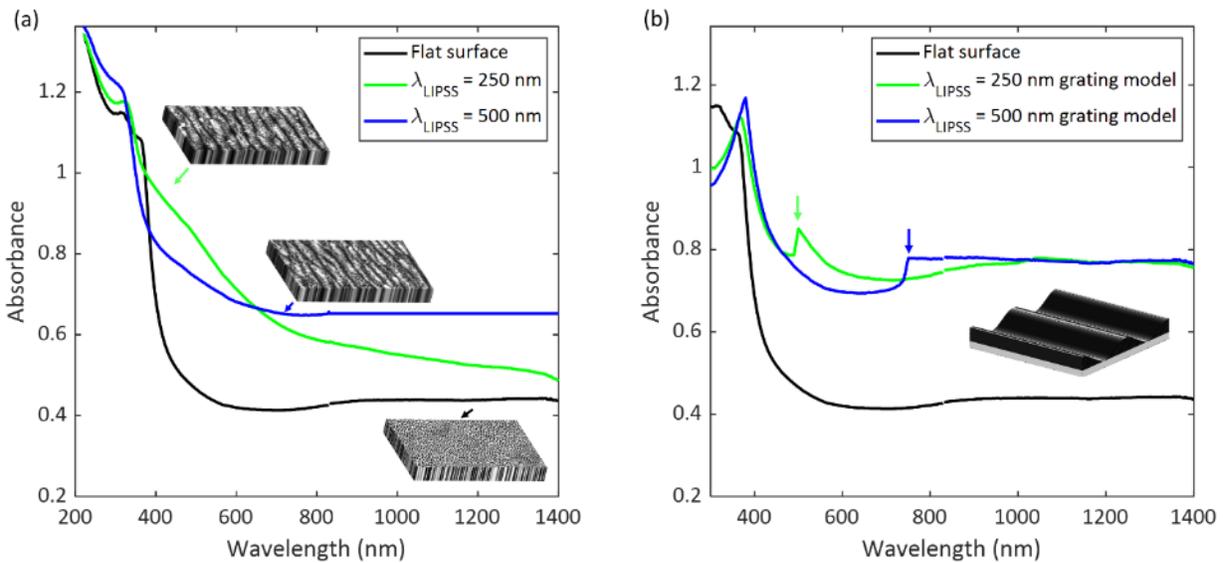

**Figure 3.** Increased visible to near-infrared light extinction of the LIPSS-Sticks: (a) Diffuse reflectance spectra for as-anodized samples (flat surface) and LIPSS-Sticks samples with initial LIPSS periodicities of $\lambda_{LIPSS}$ = 250 nm and 500 nm. Inset shows the surface morphology. (b) Simulated extinction of grating structures of TiO$_2$ with the same periodicity as the LIPSS-Sticks (expanded to 500 nm and 750 nm after anodization compared to the initial LIPSS period).

To a first approximation, we can model the increased light scattering by considering the optical scattering of a grating structure composed of uni-directional sinusoidal surface ripples. The simulated geometry has a peak-to-trough depth of 100 nm and an expanded period of 500 nm and 750 nm, due to the growth of nanotubes widening the LIPSS-Sticks's spatial period (Figure 2). The scattering of the grating was computed using GD-Calc [35], which solves Maxwell's equations using the rigorous coupled wave method [36]. The wavelength-dependent complex dielectric function of TiO$_2$ was taken from Ref [37], with a refractive index around 2.5 [38]. The average extinction efficiency (ratio of intensity absorbed and scattered to the incident intensity) was calculated for unpolarized (averaged *s*- and *p*- polarization) electromagnetic waves irradiating the surface of the grating, and up to 10 diffraction orders were computed.

The grating's predicted extinction spectrum (Figure 3b) exhibited generally increased broadband light extinction compared to the flat substrate. The increased visible light extinction between 400-680 nm of the substrate with a 250 nm LIPSS period was also reproduced. However, the grating simulation shows the presence of sharp diffraction orders, which are not seen in the measured sample due to the disordered distribution of the 100 nm nanotubes that are not included in our simplified model. Due to the roughness of the surface, the higher diffraction orders are not as distinct, and only the first diffraction order survives, which explains the higher visible wavelength extinction of the $\lambda_{LIPSS}$ = 250 nm substrate and the higher near-infrared wavelength extinction of the $\lambda_{LIPSS}$ = 500 nm substrate. This is only an approximate description, and a more complete picture of light scattering should account for the nanotube structure as well as the surface nanoparticles. Furthermore, the scattering calculations were performed for up to 10 diffraction orders, while the measurements were performed inside a $BaSO_4$ referenced integrating sphere, further introducing complications due to multiple scattering which are not easily accounted for.

**Growth morphology and dynamics of LIPSS-Sticks**

The dynamics of self-organization were explored by stopping anodization at 1, 3, 5 and 7 hours and measurement under the SEM. All LIPSS patterned surfaces show a markedly different morphology than the control or non-laser treated anodized $TiO_2$ (Figure 4). LIPSS-structuring of Ti often leaves surface Ti oxides and redeposited oxide nanoparticles, as seen in Fig 4(a), which is expected as we are in the thermochemical LIPSS regime under ambient conditions [20]. Anodized $TiO_2$ without laser treatment has an irregular nanotube structure with an average nanotube size of 120 nm from 1 to 5 hours, and a nanoforest geometry forming at 7 hours due to the etching and collapse of nanotube walls [39].

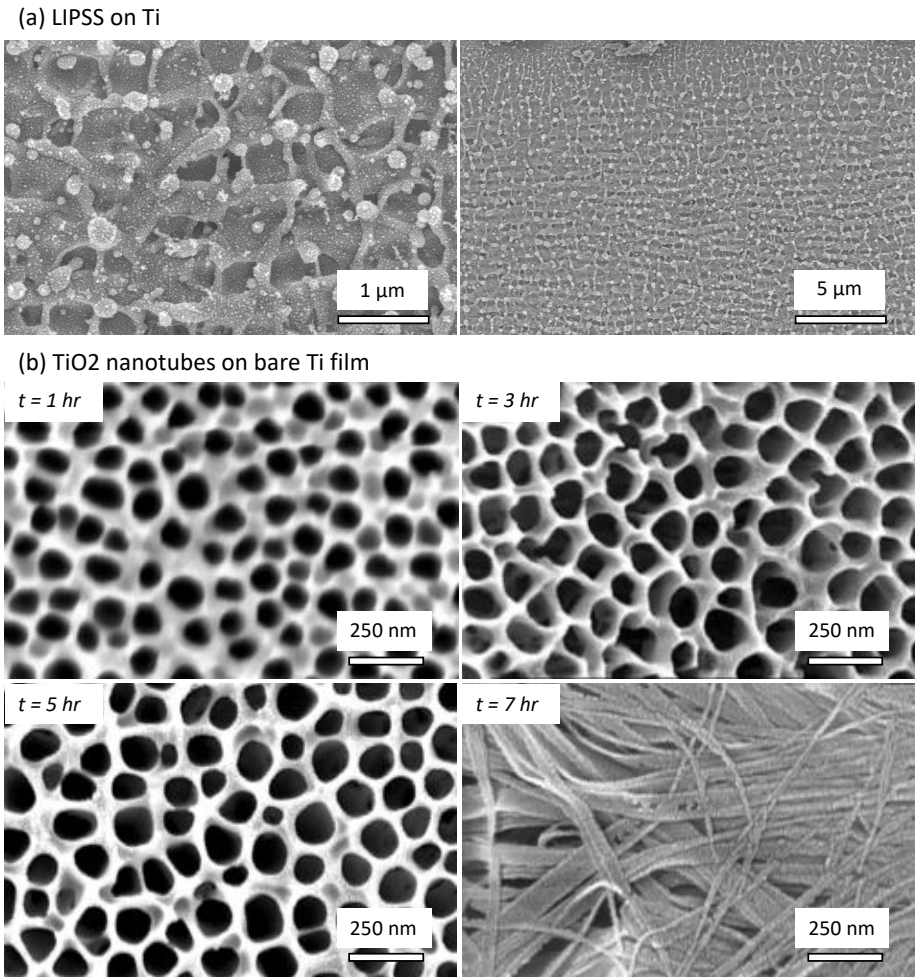

**Figure 4.** Surface morphology: (a) Laser induced periodic surface structures on Ti, and (b) Anodized $TiO_2$ surface without surface modification. Anodization time is listed.

For the LIPSS structured Ti surface anodization, the initial rippled nanostructure is preserved in the electrochemical growth. The preservation of this structure is unexpected as computational studies [40], and linear stability analysis of the oxide interface [16] suggests there is a special spatial period corresponding to the nanotube spacing that is exponentially amplified, and all other amplitudes are amplified to a lesser extent or are suppressed. Hence, the preservation of the ripples up to 5 hours of growth has to be attributed to second order effects like the pre-stress introduced to the metal film or initial oxide layer produced due to LIPSS, which will be the subject of future study.

After one hour of LIPSS-Sticks growth, the self-organized nanotube structure begins to form for both $\lambda_{LIPSS}$ = 250 nm and 500 nm surfaces (Figure 5 at high magnification, and Figure 6 at low magnification). The initial periodicity due to the LIPSS patterning is retained and results in a wave-like modulation of the nanotubes' surface profile with a larger period of 300 nm and 600 nm (white arrows in Figure 6 at T = 1 hour). The increase in spatial period is due to the growth of nanotube walls with a typical thickness of 20 nm. Nanotube diameters within the valleys of the ripple are larger than those at the hills (red circles in Figure 5). This observation can be explained by the enhanced electric field within the valleys, causing a larger valley nanotube size than at the crests. This is analysed by computational modelling in the next section. The nanotube wall thicknesses remain constant between the two patterns, which indicates that the LIPSS structures act to modify the overall surface relief without altering the shape of the nanotube walls.

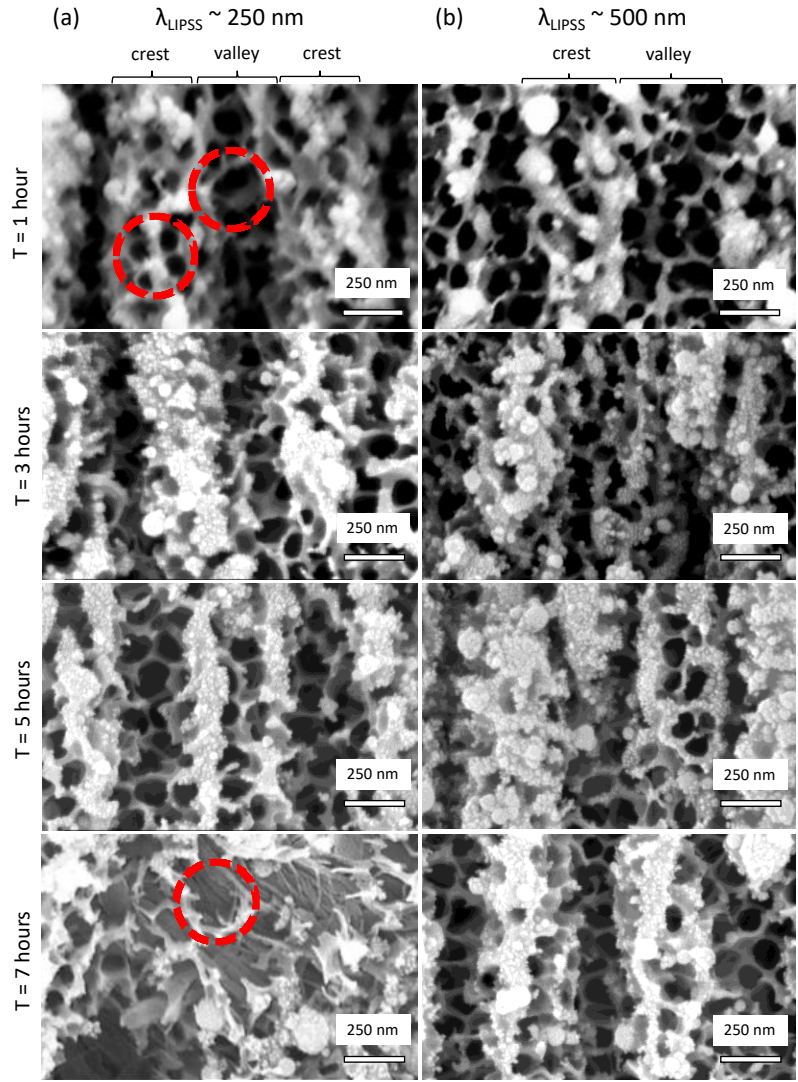

**Figure 5**. Time evolution of the surface morphology of LIPSS-Sticks with initial LIPSS periods: (a) $\lambda_{LIPSS}$ = 250 nm and (b) $\lambda_{LIPSS}$ = 500 nm. Red circles indicate the different nanotube geometries in the valleys v.s. crests of the structure, and the nano-forest formed after 7 hours of anodization.

After three hours of anodization, the nanotube walls become thicker and the oxide layer itself increases in cross-sectional thickness to ~20 μm. The periodicity remains similar to that at 1 hour anodization. Nanoparticle deposits began to grow on the surface of both samples, although more prominently in $\lambda_{LIPSS}$ = 500 nm sample, and more prominently on the crests than the valleys. We suggest that these nanoparticles are nucleated from the spheroidal ejected material created during the LIPSS patterning process (Figure 4a). These ejected particles are composed of Ti and $TiO_2$, determined via energy-dispersive X-ray spectroscopy (Supplementary Info S4). These then grow into the larger deposits seen in Figure 5. The cross-sectional structure in Figure 3 shows that the nanotube structure remains intact through the bulk of the material.

As the anodization time increases to 5 hours, the nanoparticle deposits continue to grow in size and the layer thickness increases to ~40 µm. In some areas, the nanoparticle deposits grow over the top of the nanotubes, especially in the hills of the ripples, and the nanotube walls begin to be further etched and display ribbon-like structures (Figure 2). Nanotube cross-sections are much better developed for the $\lambda_{LIPSS}$ = 500 nm sample. At 7 hours of anodization, the nano-forest geometry forms on $\lambda_{LIPSS}$ = 250 nm patterned structure, while the 500 nm patterned surface retains its nanotubes, except on the patterned structure's edges. The formation of the nanoforest is expected, as the non-laser treated Ti foil has a nano-forest geometry at 7 hours of anodization. In the cross-sectional images in Figure 2, we can see how the edges of the nanotubes become etched and collapse down to the nano-forest of nanotube walls on the surface forming a thick mat. Further cross-sectional images can be seen in Supplementary Info Section S6.

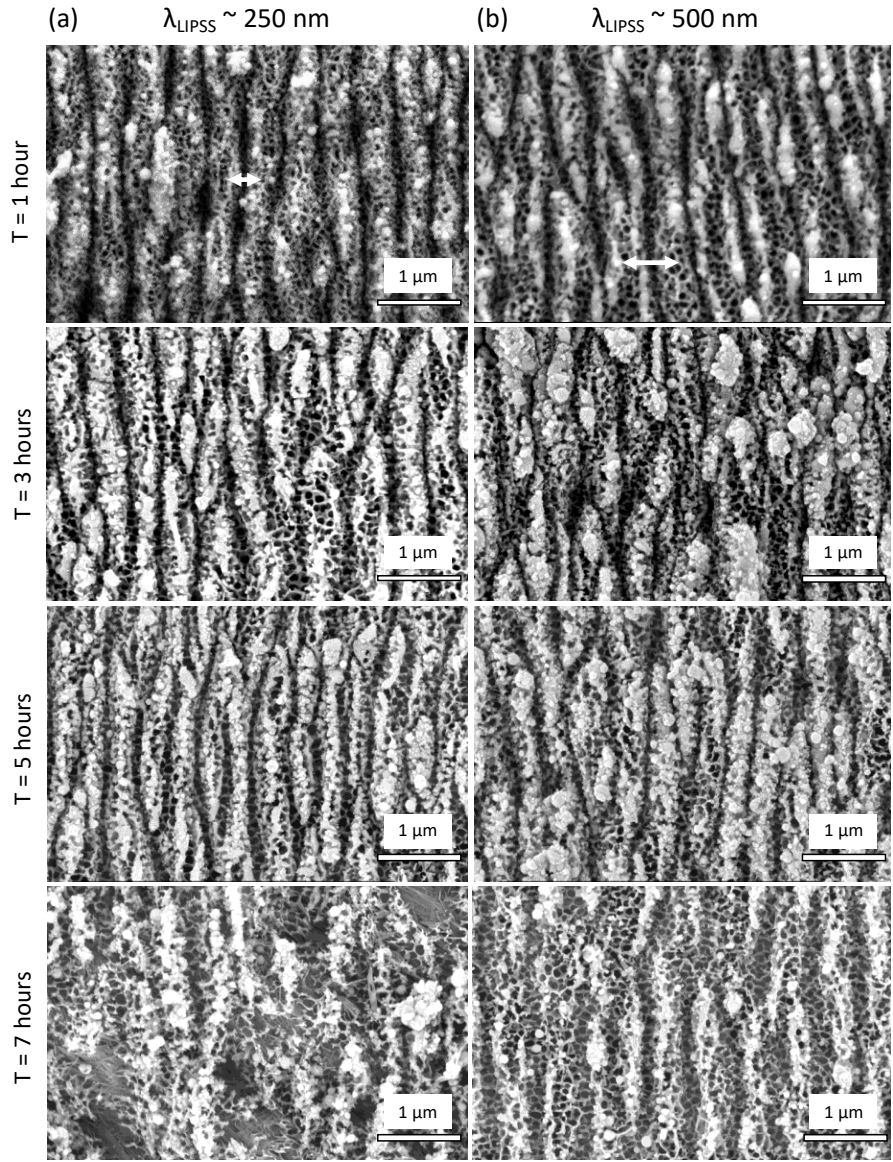

**Figure 6.** Time evolution of the surface morphology of LIPSS-Sticks at lower magnification for initial LIPSS periods: (a) $\lambda_{LIPSS}$ = 250 nm and (b) $\lambda_{LIPSS}$ = 500 nm. Arrows show the enlarged periodicity of the nanotube surface compared to the initial LIPSS spatial period.

**Electrostatic simulations of nanotube growth**

To understand the nanotube growth, we develop a simple electrostatic model simulating the electric field distribution and the current density within the LIPSS patterned Ti foil prior to and during anodization. The material and geometric parameters used are listed in Supplementary Table S2. We model the 2D simplified problem and only consider $TiO_2$ and not the electrolyte due to the higher conductivity of the electrolyte as most potential drop occurs over the insulating oxide layer [41]. The bottom metal-oxide interface has a potential equal to the applied voltage of 60 V and the top oxide-electrolyte interface is assumed to be at 0 V. Periodic boundary conditions are imposed on the edges of the oxide layer. The electrostatic field solutions for different geometries, corresponding to the evolving morphology of the nanotube array, are plotted in Figure 7.

We noted in an earlier section that the nanotube diameter in the valley of the LIPSS ripple is larger than at the edge. This can be rationalized by looking at the current density magnitude at the first stage of anodization (Figure 7a). The valley acts as the shortest path from the bottom surface at 60 V to the "ground" at 0 V, hence most of the current density passes through this region. This enhances the electric field-assisted dissolution process, deepens pores, and enlarges the pore size [42, 43]. Therefore, at a later stage of anodization, the pore that is initiated in the valley has a larger diameter and the pores near the crests have a smaller diameter (Figure 7b). However, once further pores are produced, the individual pores have a high current density and continue to deepen. The initial LIPSS ripples are preserved and not etched away, as the current density is low at the surface, thus keeping field-assisted dissolution to a minimum at the top.

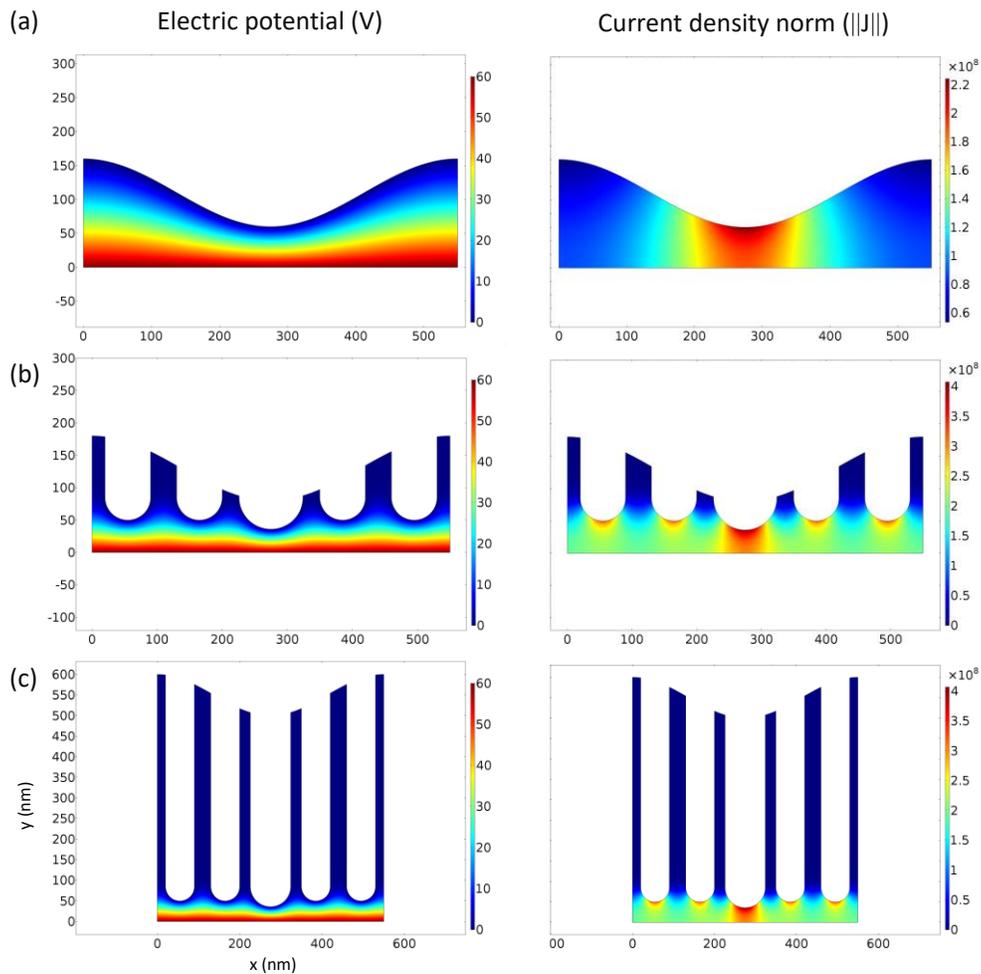

**Figure 7.** Electrochemical model of the anodization of LIPSS patterned Ti foil. The x-y axes are lengths in nm. (left:) Electric potential (V) map in V, (right:) Magnitude of current density (||J||) in Am$^{-2}$. The colour bars on the right of each plot indicate the magnitude of the corresponding

physical variable. Different stages of anodization: (a) Initial LIPSS patterned surface, (b) Formation of pores, and (c) Growth of nanotubes

Our model offers a simple explanation for the nanotube size along the LIPSS ripples, but neglects several key processes. We do not consider the diffusion of ions, the evolution of the interface due to stress and overpotential, the oxide electrochemical growth, the etching of the oxide by the fluoride electrolyte, and the initial LIPSS oxide layer formed during laser ablation. A more complete model is being developed to consider the effects of the electrolyte and surface stresses evolved in the presence of a laser templated surface. Despite that, we believe that the qualitative nature of the explanation still holds, because the electric field strength and electric currents remain the most important contributors to the initial stage of nanotube formation and pore size [43, 44].

## **Conclusions**

The novel use of laser induced periodic surface structures (LIPSS) on the surface of a metal to template the growth of photo-catalytically active metal oxide nanotube arrays onto a ripple structure is demonstrated. $TiO_2$ nanotubes are well-known photocatalysts, but have poor visible light collection efficiency. Approaches like thermal annealing introduce oxygen vacancy defects and increase the visible light absorption at the cost of reducing the electron-hole separation lifetimes. The reduced lifetime degrades the photocatalytic efficiency [13]. Ripples/grating-like nanostructures have been shown to enhance the trapping of visible light on materials. We have fabricated structures that have enhanced visible and near-infrared light extinction, which we understand to the first order as a grating scattering effect. More efficient visible light absorption from structural enhancement will reduce the amount of oxygen vacancies needed to enhance the visible light photocatalytic activity [10, 45-47]. Hence, the disadvantage of increased electron-hole recombination can be limited while still increasing the visible light collection efficiency. Furthermore, the grating structure increases the surface area of nanotubes during catalysis, potentially increasing its activity.

We also investigated the evolution of the microstructure of $TiO_2$ nanotubes and attempted to rationalize the nanotube distribution and sizes using computational modelling of the electric field during anodization. The electrostatic model, although limited, is able to explain the relative sizes of the nanotubes between the valleys of the ripple and its crests.

This is a general nanostructure fabrication method, and can be applied to a wide range of transition metal oxides. Electrochemical growth of Cu [48], Zr [49], Hf [50], Ta [51], Nb [52], W [53], Fe [54], Ti, Ti alloys (TiNb, TiZr, TiW, TiTa) [9], and Al [55] nanotube arrays have been observed, and LIPSS has been shown for Cu [56], W [57], Fe [58], Ti, and Al [59]. Hence the scope for the expansion of this technique is vast, and allows the nanostructure control of metal-oxide electro-catalysts spanning a wide range of redox potentials for different chemical reactions. We anticipate that this approach will be useful for tungsten oxides, where the native oxide's bandgap already lies in the visible regime, and tungsten oxide LIPSS-Sticks should have enhanced light collection efficiency and photocatalytic activity. Furthermore, there are other methods to produce LIPSS, including the use of circularly polarized light to create cross-hatch LIPSS structures [60]. We anticipate that the nanotube morphology by cross-hatch LIPSS templating will be different to that of linear LIPSS. A further application of the nanostructured surface would be to change the wettability of $TiO_2$ surfaces further using a hierarchy of scales [61].

**Methods and Materials**

*Laser treatment of Ti*: Prior to laser treatment, to ensure an even surface to the level of tens of nm surface roughness, Ti foil (0.2 mm thick, 99.96%) was electropolished under 40 V constant voltage in a 1:9 volume ratio perchloric acid:acetic acid. Femtosecond laser treatment (800 nm, 130 fs pulse duration, 1 kHz, Legend Elite, Coherent Inc., USA) was performed on Ti foil. The fluence was varied between 50 to 255 mJ·cm$^{-2}$ and the number of pulses overlapped at each point between 1 - 500. For templated electrochemical growth, two fluence-pulse combinations were chosen: 0.105 J·cm$^{-2}$, 250 pulses; and 0.087 J·cm$^{-2}$, 50 pulses, corresponding to laser-induced periodic structures of 250 nm and 500 nm spatial period respectively. An area of 2 cm by 2 cm was patterned on the electropolished Ti foil prior to electrochemical anodization.

*Electrochemical anodization of Ti*: The laser patterned Ti samples were anodized in an ethylene glycol electrolyte containing 0.25 wt.% $NH_4F$, 2 vol.% $H_2O$ for under 60 V constant voltage. The dimensions of the Ti foil used was 5 cm x 4 cm and the working electrode was placed 2 cm away from a Pt counter-electrode. The anodization time was varied at 1, 3, 5, and 7 hours to observe the evolution of nanotube morphology. After anodization, the samples were ultrasonicated with ethanol and distilled water and dried at 150°C to remove organic species.

*Surface characterization*: Surface morphology was observed using an SEM (JEOl JCM6000, Coherent), FE-SEM (FEI XL30 S-FEG, Phillips) and optical profilometer (Bruker Contour GTK, Bruker Inc). Ultraviolet-visible (Perkin-Elmer) reflectance spectroscopy with an integrating sphere attachment was used to measure the extinction spectrum of the sample, referenced to a $BaSO_4$ standard.

*Computational simulations and image processing*: The electromagnetic spectrum simulations were performed with the computational package GD-Calc [35], implemented in Matlab 2016a (Mathworks Inc). The code solves Maxwell's equations on a grating structure using the rigorous coupled wave approximation [36]. All image processing was done with ImageJ [62]. Electrochemical simulations were performed with COMSOL Multiphysics (Comsol Inc), using the AC/DC module.

AUTHOR INFORMATION

**Corresponding Author**

raru090@aucklanduni.ac.nz, *c.simpson@auckland.ac.nz, w.gao@auckland.ac.nz

**Present Addresses**

†NanoPhotonics Centre, Cavendish Laboratories, University of Cambridge

**Author Contributions**

RA performed the laser ablation and characterization experiments and wrote the paper with WG and MCS. JD performed the electrochemical anodization.

**Funding Sources**

Ministry of Business, Innovation and Employment Grants (MBIE) (UOAX1202, UOAX1416)

**ACKNOWLEDGMENTS**

We thank Dr. Catherine Hobbis, Dr. Patrick Conor, and Dr. Marcel Schaefer for aid in SEM measurements, Dr. Colin Doyle and Junzhi Ye for the X-ray photoelectron spectroscopy, and Dr. Wanting Chen and Prof Geoff Waterhouse for the UV-Vis reflectance spectroscopy measurements. We also thank Prof David Williams and Dr. Baptiste Auguie for discussions around the electrochemistry and optical modelling respectively.